\documentclass[12pt,epsf]{article}
\catcode`\@=11 \@addtoreset{equation}{section}  
\renewcommand{\theequation}                     
         {\arabic{section}.\arabic{equation}}   

\title{The CP(N-1) Affine Gauge Theory in the
Dynamical Space-time}
\author{P.Leifer$^1$}
\date{Bar-Ilan University,
Ramat-Gan, Israel}
\begin{document}
\maketitle \footnotetext[1]{On leave from Crimea State Engineering
and Pedagogical University, Simferopol, Crimea, Ukraine}
\begin{abstract}
An attempt to build quantum theory of field (extended) objects
without a priori space-time geometry has been represented.
Space-time coordinates are replaced by the intrinsic coordinates in
the tangent fibre bundle over complex projective Hilbert state space
$CP(N-1)$. The fate of quantum system modeled by the generalized
coherent states is rooted in this manifold. Dynamical
(state-dependent) space-time arises only at the stage of the quantum
``yes/no" measurement. The quantum measurement of the gauge ``field
shell'' of the generalized coherent state is described in terms of
the affine parallel transport of the local dynamical variables in
$CP(N-1)$.
\end{abstract}
\vskip 0.1cm
\section{Introduction}
The deep disagreement between general relativity and quantum theory
is well known \cite{Penrose}. In this work I would like represent
the model of quantum geometry, intended to reach the desirable
``peaceful coexistence" between these theories. The proposed scheme
is inherently based on the notion of the relative quantum
amplitudes. Formally, I deal with ``classical" non-linear field
theory developed over the complex projective Hilbert space
$CP(N-1)$. It is worth while to emphasize, that my approach
\cite{Le1,Le2,Le3,Le4,Le5} quite differs from number of works using
$CP(N-1)$; see for example \cite{CMP,Hughston1,Hughston2}.

The sketch of the proposed scheme is as follows:

a). I use the realization of the $G=SU(N)$ acting on the states $|S>
\in {\cal{H}}=C^N$ in terms of local dynamical variables (LDV's)
represented by the tangent vectors to $CP(N-1)$ (the operators of
differentiation).

b). Quantum measurement realized as a perturbation of the
generalized coherent quantum state (GCS).

c). The self-identification of quantum system is realized by the
affine parallel transport of its local dynamical variables, agrees
with Fubini-Study metric.

d). Variation principle applied to the local Hamiltonian vector
field leads to quasi-linear PDE field equations for the ``field
shell" of the GCS. This ``field shell" represents some ``quantum
potential" of the model extended ``particle" corresponding GCS.

e). The ``yes/no" measuring process formulated as a detection of
this extended particle serving for the establishment of the local
state-dependent space-time structure.

This approach leads to some conclusion concerning so-called
measurement problem. It is convenient to refer to the encyclopedic
book of R. Penrose \cite{Penrose}.

1. Projective postulate and null-measurement.

The so-called null-measurement  (see paragraph 22.7, \cite{Penrose})
is in fact a non-relevant construction, since the conclusion that
``we know that photon is in state $|\rho>$ even though it has not
interacted with the detector (in the transmission channel
$|\tau>$-P.L.) at all", is based on the explicit belief that the
photon already has passed splitter. But photon might be simply
absorbed even before this splitter. Therefore, strictly speaking, we
do not have reliable information without a detection in the
corresponding channel. This example shows that if one has left some
gap between two successive quantum states, the application of the
projective postulate (if no $|\tau>$, then $|\rho>$) is meaningless.

In the framework of my model the projection acts continuously and
locally along $CP(N-1)$ trajectory of GCS onto the corresponding
tangent spaces, since it is the covariant differentiation of vector
fields representing local dynamical variables (LDV's) on $CP(N-1)$.

2. Deformation of GCS during interaction used for measurement.

Let me discuss dynamics of Schr\"odinger's lump during measurement
(see paragraph 30.10, \cite{Penrose}). This construction is a humane
version of the Schr\"odinger's cat. In distinguish with so
complicated system as poisoned cat, and indefinite displaced lump of
matter, we would like discuss the deformation of GCS which is
theoretically analyzable.

First of all I should note that the assumption that ``the energy in
each case is the same" may be correct only approximately, say, in
the case of adiabatic ``kicking" of the lump. The finite time of
transition unavoidably leads to the acceleration of the lump of
matter, to the deformation of its quantum state \cite{Le6,Le7}, and
to the shift of mass-energy. Hence, the superposition state is not
stationary (beating) and, therefore, this is useless for our
decision about real interaction process of the photon and splitter
(as well as in the original ``comic example" of Schr\"odinger's cat
demonstrating incompleteness of the wave function description of an
nuclear decay \cite{Schr1}).

In the framework of my model, the GCS of the lump is ``kicked'' in
the first approximation by the coset transformations of $SU(N)$
group. The coefficient functions of the $SU(N)$ generators obey some
quasi-linear relativistic field equations in the local dynamical
space-time \cite{Le1,Le2,Le3}.

3. The difference of the masses of the original and the displaced
lumps leads to different time-like Killing vectors (if any) in the
vicinities of two lumps. This is an obstacle to write Schr\"odinger
equations for superposed wave function. But, who does need it? This
is rather a privilege than a defect, since one has a natural
decoherence mechanism.

In the framework of my model one has state-dependent space-times
arising as specific cross-section of the tangent fibre bundle over
$CP(N-1)$. Linear superposition has a sense only in dynamical
space-time in the quantum system (setup, lump) under the condition
of {\it physical integrity}. In general, the formulation of the
physical integrity is a difficult problem; in my model the GCS
expresses this property. This leads to the dynamics in the tangent
fibre bundle over $CP(N-1)$. All thought compounds of free
independent systems are trivial since they live in the tensor
product of the state spaces.

Below will be introduced some fundamental notions of my
construction.
\section{Action states with entire number N of $\hbar$}
The masses of known ``elementary'' particles $m_J$ are in the
fundamental de Broglie relation to corresponding internal
frequencies  $\omega_J$:
\begin{equation}
\frac{\omega_J}{m_J}=\frac{c^2}{\hbar}.
\end{equation}
If one treat the $U=c^2$ as the cosmic potential, then arise the
natural question about the micro-selective mechanism capable produce
very specific spectrum of frequencies. In the ordinary quantization
scheme it is assumed that the oscillator is really some fundamental
entity. But the spectrum of oscillator is equidistant and unbounded
whereas the mass-spectrum of ``elementary'' particles does not.
Furthermore, the classical soliton-like solution cannot be
decomposed into harmonic waves, hence quantum solitons are not a
compound of quantum oscillators. I try to find a dispersion law
$\Omega(P)$ (initially in the form $\Omega(X)$) as a solution of the
non-linear field equations.

There are some additional reasons for the modification of the
``second quantization'' procedure.

{\it First.} In the second quantization method one has formally
given particles whose properties are defined by some commutation
relations between creation-annihilation operators. Note, that the
commutation relations are only the simplest consequence of the
curvature of the dynamical group manifold in the vicinity of the
group's unit (in algebra). Dynamical processes require, however,
finite group transformations and, hence, the global group structure.
The main technical idea is to use vector fields over group manifold
instead of indefinite Dirac's q-numbers. This scheme therefore
looking for the dynamical nature of the creation and annihilation
processes of quantum particles.

{\it Second.} The quantum particles (energy bundles) should
gravitate. Hence, strictly speaking, their behavior cannot be
described as a linear superposition. Therefore the ordinary second
quantization method (creation-annihilation of free particles) is
merely a good approximate scheme due to the weakness of gravity.
Thereby the creation and annihilation of particles are time
consuming dynamical non-linear processes. So, linear operators of
creation and annihilation (in Dirac sense) do exist as approximate
quantities.

{\it Third.} Nobody knows how arise a quant of energy (quantum
particle). Definitely, there is an energy quantization but the
dynamical nature of this process is unknown. Avoiding the vacuum
stability problem, its self-energy, etc., we primary quantize,
however, the action, not energy. The relative (local) vacuum of some
problem is not the state with minimal energy, it is a state with an
extremal of some action functional.

 POSTULATE 1.

\noindent {\it We assume that there are elementary quantum states
(EQS) $|\hbar a>, a=0,1,...$ of abstract Planck's oscillator whose
states correspond to the quantum motions with given number of
Planck's action quanta}.

Thereby only action subject to primary quantization but the
quantization of dynamical variables such as energy, spin, etc.,
postponed to dynamical stage. Presumably there are some non-linear
field equations describing energy (frequency) distribution, whose
soliton-like solution provides the quantization of the dynamical
variables but their field carriers - ``field shell" are smeared in
dynamical space-time. Therefore, quantum ``particles'', and, hence,
their numbers should arise as some countable solutions of non-linear
wave equations. In order to establish acceptable field equation
capable intrinsically to describe all possible degrees of freedom
defreezing under intensive interaction, we should build some {\it
universal ambient Hilbert state space} $\cal{H}$. We will use {\it
the universality of the action} whose variation capable generate any
dynamical variable. Vectors of {\it action state space} $\cal{H}$ we
will call {\it action amplitude} (AA). Some of them will be EQS's of
motion corresponding to entire numbers of Planck's quanta $| \hbar
a>$. Since the action in itself does not create gravity, it is
legible to create the linear superposition of $|\hbar a>=(a!)^{-1/2}
({\hat \eta^+})^a|0>$ constituting $SU(\infty)$ multiplete of the
Planck's action quanta operator $\hat{S}=\hbar {\hat \eta^+} {\hat
\eta}$ with the spectrum $S_a=\hbar a$ in the separable Hilbert
space $\cal{H}$. The standard basis $\{|\hbar a>\}_0^{\infty}$ will
be used with the `principle' quantum number $a=0,1,2...$ assigned by
Planck's quanta counting. Generally (AA) are their coherent
superposition
\begin{eqnarray}
|G>=\sum_{a=0}^{\infty} g^a| \hbar a>.
\end{eqnarray}
may represented of the ground state - ``vacuum" of some quantum
system. In order to avoid the misleading reminiscence about
Schr\"odinger state vector, I use $|G>,|S>$ instead of $|\Psi>$.
 In fact only finite, say, $N$ EQM may be
involved. Then one may restrict $CP(\infty)$ QPS to finite
dimensional $CP(N-1)$. Hereafter I will use the indices as follows:
$0\leq a \leq N$, and $1\leq i,k,m,n,s \leq N-1$.

\section{Quantum analog of force and SU(N) factorization}
Since any ray AA has isotropy group $H=U(1)\times U(N)$, only coset
transformations $G/H=SU(N)/S[U(1) \times U(N-1)]=CP(N-1)$
effectively act in $\cal{H}$. Therefore the ray representation of
$SU(N)$ in $C^N$ and, in particular, the embedding of $H$ and $G/H$
in $G$, require a state-depending parametrization. Hence, there is a
diffeomorphism between space of the rays marked by the local
coordinates in the map
 $U_j:\{|G>,|g^j| \neq 0 \}, j>0$
\begin{eqnarray}\label{coor}
\pi^i_{(j)}=\left\{
 \matrix{\frac{g^i}{g^j} &if& 1 \leq i < j \cr
 \frac{g^{i+1}}{g^j} &if& j \leq i < N-1}
 \right\}
\end{eqnarray}
and the group manifold of the coset transformations
$G/H=SU(N)/S[U(1) \times U(N-1)]=CP(N-1)$. This diffeomorphism is
provided by the coefficient functions $\Phi^i_{\alpha}$ of the local
generators (see below and \cite{Le6,Le7}). The choice of the map
$U_j$ means, that the comparison of quantum amplitudes refers to the
amplitude with the action $\hbar j$. The breakdown of $SU(N)$
symmetry on each AA to the isotropy group $H=U(1)\times U(N-1)$
contracts full dynamics down to $CP(N-1)$. The physical
interpretation of these transformations is given by the

POSTULATE 2.

\noindent {\it Super-equivalence  principle: the unitary
transformations of the AA may be identified with the physical
unitary fields.  The coset transformation $G/H=SU(N)/S[U(1)\times
U(N-1)]=CP(N-1)$ is the quantum analog of classical force: its
action is equivalent to some physically distinguishable variation of
GCS in $CP(N-1)$}.

The $CP(N-1)$ manifold takes the place of ``classical phase space''
\cite{ZF}, since its points, corresponding to the GCS, are most
close to classical states of motion. Two interpretations may be
given for the points of $CP(N-1)$. One of them is the
``Schr\"odinger's lump" \cite{Penrose} and the second one is the
analog of the Stern-Gerlach ``filter's orientations" discussed by
Fivel \cite{Fivel}. The root content of their physical
interpretations is that one has {\it a macroscopic (i.e. space-time)
discriminator} of two quantum states. As such, they may be used as
``yes/no'' states of some two-level detector. We will use the
``Schr\"odinger's lump" interpretation. Let us assume that GCS
described by local coordinates $(\pi^1,...,\pi^{N-1})$ correspond to
the original lump, and the coordinates $(\pi^1+\delta
\pi^1,...,\pi^{N-1}+\delta \pi^{N-1})$ correspond to the displaced
lump. Such hidden coordinates of the lump gives a firm geometric
tool for the description of quantum dynamics during interaction used
for the measuring process.

Then the question that I now want to rise is as following: {\it what
``classical field'', i.e field in space-time, correspond to the
transition from the original to the displaced lump?} In other words
we would like find the ``field shell" of the lump, its space-time
shape and its dynamics. The lump's perturbations will be represented
by the ``geometric bosons'' \cite{Le8} whose frequencies are not a
priori given, but which defined by some field equations which should
established due to a new variation problem. Before its formulation,
we ought to use in fact a sophisticated differential geometric
construction in order to avoid the clash between quantum mechanics
and general relativity \cite{Penrose}.

I will assume that all ``vacua'' solutions belong to single
separable {\it projective Hilbert space} $CP(N-1)$. The vacuum
represented by GCS is merely the stationary point of some action
functional, not solution with the minimal energy. Energy will be
associated with tangent vector field to $CP(N-1)$ giving velocity of
the action variation in respect with the notion of the
Newton-Stueckelberg-Horwitz-Piron (NSHP) time \cite{H1}. Dynamical
(state-dependent) space-time will be built at any GCS and,
particulary, at the vacuum of some ``classical'' problem (see
below). Therefore Minkowskian space-time is functionally local
(state-dependent) in $CP(N-1)$ and the space-time motion dictated by
the field equations connected with two infinitesimally close GCS.
The connection between these local space-times may be physically
established by the measurement given in terms of geometry of the
base manifold $CP(N-1)$. It seems to be like the Everett's idea
about ``parallel words'', but has of course different physical
sense. Now we are evidences of the Multiverse (omnium) concept
\cite{Penrose,W1}. I think there is only one Universe but there
exists continuum of dynamical space-times each of them related to
one point of the quantum phase space $CP(N-1)$. The standard
approach, identifying Universe with space-time, is too strong
assumption from this point of view.

\section{LDV's and tangent fibre bundles}
The state space ${\cal H}$ of the field configurations with finite
action quanta is a stationary construction. We introduce dynamics
{\it by the velocities of the GCS variation} representing some
``elementary excitations'' (quantum particles). Their dynamics is
specified by the Hamiltonian, giving time variation velocities of
the action quantum numbers in different directions of the tangent
Hilbert space $T_{(\pi^1,...,\pi^{N-1})} CP(N-1)$ where takes place
the ordinary linear quantum scheme. The temp of the action variation
gives the energy of the ``particles''.

The local dynamical variables corresponding internal symmetries of
the GCS and their breakdown should be expressed now in terms of the
local coordinates $\pi^k$. The Fubini-Study metric
\begin{equation}
G_{ik^*} = [(1+ \sum |\pi^s|^2) \delta_{ik}- \pi^{i^*} \pi^k](1+
\sum |\pi^s|^2)^{-2} \label{FS}
\end{equation}
and the affine connection
\begin{eqnarray}
\Gamma^i_{mn} = \frac{1}{2}G^{ip^*} (\frac{\partial
G_{mp^*}}{\partial \pi^n} + \frac{\partial G_{p^*n}}{\partial
\pi^m}) = -  \frac{\delta^i_m \pi^{n^*} + \delta^i_n \pi^{m^*}}{1+
\sum |\pi^s|^2} \label{Gamma}
\end{eqnarray}
in these coordinates will be used. Hence the internal dynamical
variables and their norms should be state-dependent, i.e. local in
the state space \cite{Le6,Le7}. These local dynamical variables
realize a non-linear representation of the unitary global $SU(N)$
group in the Hilbert state space $C^N$. Namely, $N^2-1$ generators
of $G = SU(N)$ may be divided in accordance with Cartan
decomposition: $[B,B] \in H, [B,H] \in B, [B,B] \in H$. The
$(N-1)^2$ generators
\begin{eqnarray} \Phi_h^i \frac{\partial}{\partial \pi^i}+c.c. \in
H,\quad 1 \le h \le (N-1)^2
\end{eqnarray}
 of the isotropy group $H = U(1)\times
U(N-1)$ of the ray (Cartan sub-algebra) and $2(N-1)$ generators
\begin{eqnarray}
\Phi_b^i \frac{\partial}{\partial \pi^i} + c.c. \in B, \quad 1 \le b
\le 2(N-1)
\end{eqnarray}
are the coset $G/H = SU(N)/S[U(1) \times U(N-1)]$ generators
realizing the breakdown of the $G = SU(N)$ symmetry of the GCS.
Furthermore, $(N-1)^2$ generators of the Cartan sub-algebra may be
divided into the two sets of operators: $1 \le c \le N-1$ (where
$N-1$ is the rank of $Alg SU(N)$) Abelian operators, and $1 \le q
\le (N-1)(N-2)$ non-Abelian operators corresponding to the
non-commutative part of the Cartan sub-algebra of the isotropy
(gauge) group. Here $\Phi^i_{\sigma}, \quad 1 \le \sigma \le N^2-1 $
are the coefficient functions of the generators of the non-linear
$SU(N)$ realization. They give the infinitesimal shift of
$i$-component of the coherent state driven by the $\sigma$-component
of the unitary multipole field rotating the generators of $Alg
SU(N)$ and they are defined as follows:
\begin{equation}
\Phi_{\sigma}^i = \lim_{\epsilon \to 0} \epsilon^{-1}
\biggl\{\frac{[\exp(i\epsilon \lambda_{\sigma})]_m^i g^m}{[\exp(i
\epsilon \lambda_{\sigma})]_m^j g^m }-\frac{g^i}{g^j} \biggr\}=
\lim_{\epsilon \to 0} \epsilon^{-1} \{ \pi^i(\epsilon
\lambda_{\sigma}) -\pi^i \},
\end{equation}
\cite{Le1}.
Then the sum of $N^2-1$ the energies of the `elementary
systems' (particle plus fields) is equal to the excitation energy of
the GCS, and the local Hamiltonian $\vec{H}$ is linear against the
partial derivatives $\frac{\partial }{\partial \pi^i} = \frac{1}{2}
(\frac{\partial }{\partial \Re{\pi^i}} - i \frac{\partial }{\partial
\Im{\pi^i}})$ and $\frac{\partial }{\partial \pi^{*i}} = \frac{1}{2}
(\frac{\partial }{\partial \Re{\pi^i}} + i \frac{\partial }{\partial
\Im{\pi^i}})$, i.e. it is the tangent vector to $CP(N-1)$
\begin{eqnarray}
\vec{H}=\vec{T_c}+\vec{T_q} +\vec{V_b} = \hbar \Omega^c \Phi_c^i
\frac{\partial }{\partial \pi^i} + \hbar \Omega^q \Phi_q^i
\frac{\partial }{\partial \pi^i} + \hbar \Omega^b \Phi_b^i
\frac{\partial }{\partial \pi^i} + c.c.. \label{field}
\end{eqnarray}
The characteristic equations for the PDE $\vec{H}|E>=E|E>$ give the
parametric representations of their solutions in $CP(N-1)$. The
parameter $\tau$ in these equations I will identify with ``universal
time of evolution'' of Newton-Stueckelberg-Horwitz-Piron-(NSHP)
\cite{H1}. This time is the measure of the GCS variation, i.e. it is
a state-dependent measure of the distance in $CP(N-1)$ (an evolution
trajectory length in the Fubini-Study metric) expressed in time
units. The energy quantization will be discussed elsewhere.

\section{Lorentz transformations and dynamical space-time }
The Einstein's analysis of the Galileo-Newtonian kinematics in an
inertial reference frame, based on the classical Maxwell's
electromagnetic field theory, led us to a new relativistic
kinematics \cite{Einstein1}. Unfortunately, similar analysis based
on the quantum theory is in a very preliminary state
\cite{Penrose,Ashtekar}. The continuation of such a work is
necessary.

It is clear that the {\it coincidence} of the ``arrow'' with some
number on the ``limb'' is in fact the coincidence of the two
space-time points. But one has in the quantum area literally neither
``arrow'' nor the ``limb''; some ``clouds'' or `` field shell" one
has instead. Thereby the uncertainty principe puts the limit for the
exact coincidence of two events. Therefore, in comparison with us,
Einstein had two privileges: he had intuitively clear classical
measuring devises (clocks, scales, rods, etc.) and the intuitively
clear spatial coincidence of two ``points'', say, the end of a rod
and the end of the scale. Without these ingredients it is difficult
to image the measurement process and even the space-time notion
itself. Generally, space-time coordinates lose direct physical sense
even in the framework of general relativity \cite{Einstein2}.
Quantum theory poses a new problem concerning operational sense of
the microscopic invariance of the space-time scale. Indeed, all
abstract (notional) tools of macroscopic laboratory (clocks, scales,
rods, etc.) one should change for the microscopic ones. Note, Bohr's
proposal about ``classical apparatus'' is unacceptable since it is
inconsistent. We should construct now the space-time notion in the
internal quantum terms.

The notion of physical space is based on the abstraction of
separation between two material points assuming that they may be as
far as we need. This separation may be measured by some hard scale.
The ``hard scale'' in fact means the ``the same'' or identical
scale, i.e. the scale with invariant length relative some
transformation group. But in quantum theory it is inconsistent (or
it is at least questionable) to use a priori space-time symmetries.
Similar arguments is applicable to time separation because of the
specific problem of the ``time-of-arrival'' \cite{A1}. Generally
speaking, {\it space-time separation is state-dependent}. In such
situation, one should decide on what is the criterion of identity is
physically acceptable in our case. If, say, electrons used for the
measurement of separation between a source $S_1$ and a detector
$D_1$, than the most reasonable to use the criterion of the ``same
electron'' (emitted from $S_1$ and detected in $D_1$). All quantum
electrons are of course identical but there are momentum and spin
which distinguish one electron from another. But in general this
criterion is not so good as we need since we cannot be sure that the
electron detected in $D_2$ is same as in $D_1$, or even that this
has some causal connection with the previous stage of the
measurement. There is at least one reason for this verdict: the
detection of some accidental electron, e.g. due to a quantum
fluctuations, etc. Nevertheless, in the bubble chamber one may be
sure that whole visible trace belongs to ``same electron''.
Therefore, if interaction is not so drastic or, if one takes into
account all possible decay channels of an unitary multiplete, we
could formulate the criterion of identity. Let me to formulate this
criterion previously with promise to decode all components of the
statement: {\it the local Hamiltonian should be parallel transported
during a ``smooth'' evolution}. I introduce the concept of
``dynamical space-time'' as a new construction capable to detect the
coincidences of the qubit components in the formal two-level
``detector'' which is a part of the full quantum configuration
(setup modeled by the GCS). The ``extraction'' of this ``detector''
is of course more or less a free choice of an observer. It is
important only that the chosen LDV should be invariantly connected
with the qubit coherent state in respect with a one of the points of
the LDV spectrum. I will assume that the spectrum of the LDV is
known even if it is really problematic like, for example, in the PDE
eigen-problem $\vec{H}|E>=E|E>$ mentioned above.
\subsection{Embedding ``Hilbert (quantum)
dynamics" in space-time} If we would like to have some embedding of
the ``Hilbert (quantum) dynamics" in space-time we should to
formalize the quantum observation (or measurement of some internal
dynamical variable).

Mentioned above diffeomorphism between rays of $CP(N-1)$ and $SU(N)$
generators will be realized in terms of the local $SL(2,C)$ action
onto the qubit states space $C^2$ as follows.

The basis of these spaces form two vectors: the normal vector $|N>$
to the ``vacuum landscape'' $CP(N-1)$ corresponding to eigenvalue
$\lambda_D$ of measuring dynamical variable $\hat{D}$ and the
tangent vector $|T>$, generated by the coset generators of $G/H$.
The last ones describing the interaction used for the measurement
process. It is important to understand that the measurement i.e.
comparison of the expected qubit spinor $(\alpha_0,\beta_0)$ at and
measured qubit spinor $(\alpha_1,\beta_1)$ pave the way to embedding
Hilbert space dynamics into the local dynamical space-time. This is
the replacement of the notorious ``arrow'' of the measuring device,
namely: one has two-level system (logical spin $1/2$ \cite{Le6})
created by the quantum question-unitary projector onto one of the
two states $|N>,|T>$. Their coherent states are given by the qubit
spinors $(\alpha,\beta)$ being connected with infinitesimal
$SL(2,C)$ transformations give rise to the variation of the
space-time coordinates generated by local infinitesimal Lorentz
transformations. Why we can do this conclusion?

Causal classical events lie (in good approximation) on a light cone
which is invariant relative the Lorentz group. On the other hand the
formal ``Lorentz spin transformations matrix'' transform the spinor
of the quantum question being applied to measurement of some LDV
helping us to detect some event. The classical detection of an event
is based on the coincidence of the two spinors one of which
corresponds to the expectation value and the second to detecting
value of LDV. This is possible only under the tuning of orientation
by rotation and the tuning of velocity by acceleration. Therefore we
should identify ``Lorentz spin transformations matrix''  of the
qubit spinors with Lorentz transformation of classical inertial
frame.

The specific components of LDV's (see below) take the place of these
entities.  But now LDV is vector field defined over $CP(N-1)$ and
the comparison of LDV at different setups (initial and perturbed due
to interaction used for measurement) require some procedure of the
{\it self-identification}. It is impossible to compare expected and
measured LDV ``directly'' (decoherence due to $CP(N-1)$ geometry
\cite{Le4}). The affine parallel transport is quite acceptable for
this aim. The parallel transport forms the condition for the
coefficient functions of the LDV leading to the nonlinear field
equations in the local dynamical space-time.

\subsection{Differential geometry of the measuring procedure}
The measurement, i.e. attributing a number to some dynamical
variable or observable has in physics subjective as well as
objective sense. Namely: the numeric value of some observable
depends as a rule on a setup (the character of  motion of
laboratory, type of the measuring device, field strength, etc.).
However the relationships between numeric values of dynamical
variables and numeric characteristics of laboratory motion, field
strength, etc., should be formulated as invariant, since they
reflect the objective character of the physical interaction used in
the measurement process. The numbers obtained  due to the
measurements carry information which does not exist a priori, i.e.
before the measurement process. But the information comprised of
subjective as well as objective invariant part reflects the physics
of interaction. The last is one of the main topics of QFT. Since
each measurement reducible (even if it is unconscious) to the answer
of the question ``yes'' or ``no'', it is possible to introduce
formally a quantum dynamical variable ``logical spin 1/2" \cite{Le6}
whose coherent states represent the quantum bit of information
``qubit".

POSTULATE 3

{\it We assume that the invariant i.e. physically essential part of
information represented by the coherent states of the ``logical spin
1/2" is related to the space-time structure.}

Such assumption is based on the observation that on one side the
space-time is the manifold of points modeling different physical
systems (stars, atoms, electrons, etc.) artificially depleted of all
physical characteristics (material points without reference to
masses). In principle arbitrary local coordinates may be attributed
to these points. But as we know from general relativity the metric
structure depends on the matter distribution and the zero
approximation of the metric tensor $g_{\mu \nu}=\eta_{\mu \nu}+...$
gives the Lorentz invariant interval \cite{Einstein2}. On the other
hand the spinor structure of the Lorentz transformations represents
the transformations of the coherent states of the ``logical spin
1/2" or ``qubit". Thereby we can assume the measurement of the
quantum dynamical variables expressed by the ``qubit" spinor
``creates'' the local space-time coordinates. We will formulate
non-linear field equations in this local space-time due to a
variational principle referring to the generator of the quantum
state deformation.

The internal hidden dynamics of the quantum configuration given by
GCS should be somehow reflected in physical space-time. Therefore we
should solve the ``inverse representation problem'': to find locally
unitary representation of dynamical group $SU(N)$ in the dynamical
space-time where acts the induced realization of the coherence group
$SU(2)$ of the qubit spinor \cite{Le1,Le2}. Its components subjected
to the ``Lorentz spin matrix transformations'' \cite{G}. We should
build the local spinor basis invariantly related to the ground
states manifold $CP(N-1)$. First of all we have to have the local
reference frame (LRF) as some analog of the ``representation'' of
$SU(N)$. Each LRF and, hence, $SU(N)$ ``representation'' may be
marked by the local coordinates (\ref{coor}) of the ``vacuum
landscape''. Now we should almost literally repeat differential
geometry of a smooth manifold embedded in flat ambient Hilbert
space${\cal{H}}=C^N$. The geometry of this smooth manifold is the
projective Hilbert space equipped with the Fubini-Study metric
(\ref{FS}) and with the affine connection (\ref{Gamma}).

In order to express the measurement of the ``particle's field'' in
the geometrically intrinsic terms, I assume that GCS is expressed in
the local coordinates
\begin{eqnarray}
|G(\pi^1,...,\pi^{N-1})>=(g^0(\pi^1,...,\pi^{N-1}),g^1(\pi^1,...,\pi^{N-1}),
...,g^{N-1}(\pi^1,...,\pi^{N-1}))^T,
\end{eqnarray}
where $\sum_{a=0}^N |g^a|^2= R^2$, and, hence,
\begin{eqnarray}
g^0(\pi^1,...,\pi^{N-1})=\frac{R^2}{\sqrt{R^2+\sum_{s=1}^{N-1}|\pi^s|^2}},
\end{eqnarray} and for $1\leq i\leq N-1$ one has
\begin{eqnarray}
g^i(\pi^1,...,\pi^{N-1})=\frac{R
\pi^i}{\sqrt{R^2+\sum_{s=1}^{N-1}|\pi^s|^2}},
\end{eqnarray}
i.e. $CP(N-1)$ will be embedded in the Hilbert space of Planck's
quanta ${\cal{H}}=C^N$.

Then the velocity of ground state evolution relative NSHP time is
given by the formula
\begin{eqnarray}\label{41}
|H> = \frac{d|G>}{d\tau}=\frac{\partial g^a}{\partial
\pi^i}\frac{d\pi^i}{d\tau}|a\hbar>=|T_i>\frac{d\pi^i}{d\tau}=H^i|T_i>,
\end{eqnarray}
 is the tangent vector to the evolution curve
$\pi^i=\pi^i(\tau)$, where
\begin{eqnarray}\label{42}
|T_i> = \frac{\partial g^a}{\partial \pi^i}|a\hbar>=T^a_i|a\hbar>.
\end{eqnarray}
Then the ``acceleration'' is as follows
\begin{eqnarray}\label{43}
|A> =
\frac{d^2|G>}{d\tau^2}=|g_{ik}>\frac{d\pi^i}{d\tau}\frac{d\pi^k}{d\tau}
+|T_i>\frac{d^2\pi^i}{d\tau^2}=|N_{ik}>\frac{d\pi^i}{d\tau}\frac{d\pi^k}{d\tau}\cr
+(\frac{d^2\pi^s}{d\tau^2}+\Gamma_{ik}^s
\frac{d\pi^i}{d\tau}\frac{d\pi^k}{d\tau})|T_s>,
\end{eqnarray}
where
\begin{eqnarray}\label{44}
|g_{ik}>=\frac{\partial^2 g^a}{\partial \pi^i \partial \pi^k}
|a\hbar>=|N_{ik}>+\Gamma_{ik}^s|T_s>
\end{eqnarray}
and the state
\begin{eqnarray}\label{45}
|N> = N^a|a\hbar>=(\frac{\partial^2 g^a}{\partial \pi^i \partial
\pi^k}-\Gamma_{ik}^s \frac{\partial g^a}{\partial \pi^s})
\frac{d\pi^i}{d\tau}\frac{d\pi^k}{d\tau}|a\hbar>
\end{eqnarray}
is the normal to the ``hypersurface'' of the ground states. Then the
minimization of this ``acceleration'' under the transition from
point $\tau$ to $\tau+d\tau$ may be achieved by the annihilation of
the tangential component
\begin{equation}
(\frac{d^2\pi^s}{d\tau^2}+\Gamma_{ik}^s
\frac{d\pi^i}{d\tau}\frac{d\pi^k}{d\tau})|T_s>=0
\end{equation}
i.e. under the condition of the affine parallel transport of the
Hamiltonian vector field
\begin{equation}\label{par_tr}
dH^s +\Gamma^s_{ik}H^id\pi^k =0.
\end{equation}

The Gauss-Codazzi equations
\begin{eqnarray}\label{46}
\frac{\partial N^a}{\partial \pi^i}=B^s_i T^a_s \cr \frac{\partial
T_k^a}{\partial \pi^i}-\Gamma^s_{ik}T^a_s=B_{ik}N^a
\end{eqnarray}
I used here instead of the anthropic principle \cite{Penrose,W1}.
These give us dynamics of the vacuum (normal) vector and the tangent
vectors, i.e. one has the LRF dynamics modeling the ``moving
representation'' or moving quantum setup
\begin{eqnarray}\label{47}
\frac{d N^a}{d \tau}=\frac{\partial N^a}{\partial \pi^i} \frac{d
\pi^i}{d\tau}+c.c.= B^s_i T^a_s \frac{d \pi^i}{d\tau} +c.c. = B^s_i
T^a_s H^i +c.c.; \cr \frac{d T_k^a}{d \tau}=\frac{\partial
T_k^a}{\partial \pi^i}\frac{d \pi^i}{d\tau} +c.c. =
  (B_{ik}N^a+\Gamma^s_{ik}T^a_s)\frac{d \pi^i}{d\tau}+c.c. \cr
= (B_{ik}N^a+\Gamma^s_{ik}T^a_s) H^i+c.c.
\end{eqnarray}
Please, remember that $0 \leq a \leq N$, but $1\leq i,k,m,n,s \leq
N-1$. The tensor $B_{ik}$ of the second quadratic form of the ground
states ``hypersurface'' is as follows:
\begin{eqnarray}\label{48}
B_{ik} =<N|\frac{\partial^2 |G>}{\partial \pi^i \partial \pi^k}.
\end{eqnarray}

Now one should build the qubit spinor in the local basis $(|N>,|D>)$
for the quantum question in respect with the measurement of some
local dynamical variable $\vec{D}$ at some GCS which may be marked
by the normal $|N>$. We will assume that there is {\it natural state
$\widetilde{|D>}$ of the quantum system in the LRF representation}
equal to the renormalized lift of LDV $\vec{D} \in T_{\pi}CP(N-1)$
into the environmental Hilbert space $\cal{H}$, and there is {\it
expectation state} $\hat{D}|D_{expect}>=\lambda_D|D_{expect}>$,
associated with the tuning of ``measuring device''. This notional
measuring device is associate with the local unitary projector along
the normal $|N>$ onto the natural state $\widetilde{|D>}$. In fact
it defines the covariant derivative in $CP(N-1)$. The lift-vectors
$|N>,|D>$ are given by the solutions of (\ref{47}) arising under
interaction used for the measurement of the LDV $\vec{D}$. In
general $|D>$ it is not a tangent vector to $CP(N-1)$. But
renormalized vector defined as the covariant derivative
$|\widetilde{D}>=|D>-<Norm|D>|Norm>$ is a tangent vector to
$CP(N-1)$ if $|Norm>=\frac{|N>}{\sqrt{<N|N>}}$. The operation of the
$|\widetilde{D}>$ renormalization is the orthogonal (unitary)
projector. Indeed,
\begin{eqnarray}
\widetilde{|\widetilde{D}>}= \widetilde{(|D>-<Norm|D>|Norm>)}\cr =
|D>-<Norm|D>|Norm> \cr - <Norm|(|D>-<Norm|D>|Norm>)|Norm> \cr
=|D>-<Norm|D>|Norm> = |\widetilde{D}>.
\end{eqnarray}
Then at the point $(\pi^1,...,\pi^{N-1})$ one has two components of
the qubit spinor
\begin{eqnarray}\label{513}
\alpha_{(\pi^1,...,\pi^{N-1})}=\frac{<N|D_{expect}>}{<N|N>} \cr
\beta_{(\pi^1,...,\pi^{N-1})}=\frac{<\widetilde{D}|D_{expect}>}
{<\widetilde{D}|\widetilde{D}>}
\end{eqnarray}
then at the infinitesimally close point
$(\pi^1+\delta^1,...,\pi^{N-1}+\delta^{N-1})$ one has new qubit
spinor
\begin{eqnarray}\label{514}
\alpha_{(\pi^1+\delta^1,...,\pi^{N-1}+\delta^{N-1})}=\frac{<N'|D_{expect}>}
{<N'|N'>} \cr \beta_{(\pi^1+\delta^1,...,\pi^{N-1}+\delta^{N-1})}=
\frac{<\widetilde{D}'|D_{expect}>}{<\widetilde{D}'|\widetilde{D}'>}
\end{eqnarray}
where the basis $(|N'>,|\widetilde{D}'>)$ is the lift of the
parallel transported $(|N>,|\widetilde{D}>)$ from the
infinitesimally close $(\pi^1+\delta^1,...,\pi^{N-1}+\delta^{N-1})$
back to $(\pi^1,...,\pi^{N-1})$.

These two infinitesimally close qubit spinors being expressed as
functions of $\theta,\phi,\psi,R$ and
$\theta+\epsilon_1,\phi+\epsilon_2,\psi+\epsilon_3,R+\epsilon_4,$
represented as follows
\begin{eqnarray}\label{s1}
\eta = R \left( \begin {array}{c} \cos \frac{\theta}{2}(\cos
\frac{\phi_1- \psi_1}{2} - i\sin \frac{\phi_1 - \psi_1}{2}) \cr \sin
\frac{\theta}{2} (\cos \frac{\phi_1+\psi_1}{2} +i \sin
 \frac{\phi_1+\psi_1}{2})  \end {array}
 \right)
 = R\left( \begin {array}{c} C(c-is) \cr S( c_1+is_1)
\end
{array} \right)
\end{eqnarray}
and
\begin{eqnarray}
\eta+\delta \eta = R\left( \begin {array}{c} C(c-is) \cr S(
c_1+is_1) \end {array} \right) \cr + R\left( \begin {array}{c}
S(is-c)\epsilon_1-C(s+i c)\epsilon_2+
C(s+ic)\epsilon_3+C(c-is)\frac{\epsilon_4}{R} \cr
 C(c_1+is_1)\epsilon_1+S(ic_1-s_1)\epsilon_2-S(s_1-ic_1)\epsilon_3
+S(c_1+is_1)\frac{\epsilon_4}{R}
\end
{array}
 \right)
\end{eqnarray}
may be connected with infinitesimal ``Lorentz spin transformations
matrix'' \cite{G}
\begin{eqnarray}
L=\left( \begin {array}{cc} 1-\frac{i}{2}\tau ( \omega_3+ia_3 )
&-\frac{i}{2}\tau ( \omega_1+ia_1 -i ( \omega_2+ia_2)) \cr
-\frac{i}{2}\tau
 ( \omega_1+ia_1+i ( \omega_2+ia_2))
 &1-\frac{i}{2}\tau( -\omega_3-ia_3)
\end {array} \right).
\end{eqnarray}
Then accelerations $a_1,a_2,a_3$ and angle velocities $\omega_1,
\omega_2, \omega_3$ may be found in the linear approximation from
the equation
\begin{eqnarray}\label{equ}
\eta+\delta \eta = L \eta
\end{eqnarray}
as functions of the qubit spinors components depending on local
coordinates $(\pi^1,...,\pi^{N-1})$.

Hence the infinitesimal Lorentz transformations define small
``space-time'' coordinates variations. It is convenient to take
Lorentz transformations in the following form $ct'=ct+(\vec{x}
\vec{a}) d\tau, \quad \vec{x'}=\vec{x}+ct\vec{a} d\tau
+(\vec{\omega} \times \vec{x}) d\tau$, where I put
$\vec{a}=(a_1/c,a_2/c,a_3/c), \quad
\vec{\omega}=(\omega_1,\omega_2,\omega_3)$ \cite{G} in order to have
for $\tau$ the physical dimension of time. The coordinates $x^\mu$
of points in this space-time serve in fact merely for the
parametrization of deformations of the ``field shell'' arising under
its motion according to non-linear field equations \cite{Le1,Le2}.

\section{Field shell equations (FSE)}
In order to find the ``field shell'' of the perturbed GCS one should
establish some wave equations in the dynamical space-time. All these
notions require more precise definitions. Namely, say, in the
simplest case of $CP(1)$, the ``field shells'' being represented in
the spherical coordinates are the classical vector fields
$\Omega^{\alpha}=\frac{x^{\alpha}}{r}(\omega +i \gamma), \quad 1\leq
\alpha \leq 3 $ giving the temps of the GCS variations. The tensor
fields $1\leq \alpha \leq 8,15,...,N^2-1$ will be discussed
elsewhere. Note, that the maximal number of EQS $a=0,1,...N,...$ now
strongly connected with the tensor character of the GCS driving
field $\Omega^{\alpha}$. These fields are ``classical'' since they
are not subjected to quantization directly, i.e. by the attribution
of the fermionic or bosonic commutation relations. They obey to
nonlinear field equations. Their internal dynamical variables like
spin, charge, etc.,  will be a consequence of their dynamical
structure.

``Particle'' now associated with the ``field shell'' in the
dynamical space-time (see below), given locally by the Hamiltonian
vector field $\vec{H}$. At each point $(\pi^1,...,\pi^{N-1})$ of the
$CP(N-1)$ one has an ``expectation value'' of the $\vec{H}$ defined
by a measuring device. But displaced GCS may by reached along of one
of continuum pathes. Therefore the comparison of two vector fields
and their ``expectation values'' in neighborhood points requires
some natural rule. The ``natural'' in our case means that the
comparison has sense only for same ``particle'' or for its ``field
shell''. For this reason one should have a ``self-identification''
procedure. The affine parallel transport in $CP(N-1)$ of vector
fields is a natural and the simplest rule for the comparison of
corresponding ``field shells''. Physically the self-identification
of ``particle'' literally means that its Hamiltonian vector field is
the Fubini-Study covariant constant.

But there are questions: what should coincide, and what is the
``expected'' and what is ``the detected particles'', because we have
not particles at all? Since we have only the unitary fields
$\Omega^{\alpha}$ as parameters of GCS transformations we assume
that in accordance with the super-equivalence principle under the
infinitesimal shift of the unitary field $\delta \Omega^{\alpha}$ in
the dynamical space-time, the shifted Hamiltonian field should
coincide with the infinitesimal shift of tangent Hamiltonian field
generated by the parallel transport in $CP(N-1)$  during NSHP time
$\delta \tau$ \cite{H1}. Thus one has
\begin{equation}
\hbar (\Omega^{\alpha} + \delta \Omega^{\alpha} ) \Phi^k_{\alpha}=
\hbar \Omega^{\alpha}( \Phi^k_{\alpha} - \Gamma^k_{mn}
\Phi^m_{\alpha} V^n \delta \tau)
\end{equation}
 and, hence,
\begin{equation}
\frac{ \delta \Omega^{\alpha}}{\delta \tau} = -
\Omega^{\alpha}\Gamma^m_{mn} V^n
\end{equation}.

We introduce the dynamical space-time coordinates $x^{\mu}$ as
state-dependent quantities, transforming in accordance with the
local Lorentz transformations $x^{\mu} + \delta x^{\mu} =
(\delta^{\mu}_{\nu} + \Lambda^{\mu}_{\nu} \delta \tau )x^{\nu}$. The
parameters of $\Lambda^{\mu}_{\nu} (\pi^1,...,\pi^{N-1})$ depend on
the local transformations of LRF in $CP(N-1)$ described in the
previous paragraph. Assuming a spherically symmetrical solution, we
will use the coordinates $(x^0=ct,x^1=r\sin \Theta \cos \Phi,
x^2=r\sin \Theta \sin \Phi, x^3=r\cos \Theta)$. In the case of
spherical symmetry, $\Omega^1=(\omega+i \gamma) \sin \Theta \cos
\Phi, \Omega^2=(\omega+i \gamma) \sin \Theta \sin \Phi,
\Omega^3=(\omega+i \gamma) \cos \Theta)$ and in the general case of
the separability of the angle and radial parts, one has
$\Omega^{\alpha}=\sum C_{l,m}^{\alpha} Y_{l,m}(\Theta,\Phi)(\omega+i
\gamma)=\sum C_{l,m}^{\alpha} Y_{l,m}(\Theta,\Phi)\Omega$. Then
taking into account the expressions for the ``4-velocity"
$v^{\mu}=\frac{\delta x^{\mu}}{\delta \tau} = \Lambda^{\mu}_{\nu}
(\pi^1,...,\pi^{N-1})
 x^{\nu} $ one has the field equation
\begin{equation}\label{FSE}
v^{\mu} \frac{\partial \Omega}{\partial x^{\mu} } = -
\Omega\Gamma^m_{mn} V^n,
\end{equation}
where
\begin{equation}
\matrix{ v^0&=&(\vec{x} \vec{a}) \cr
 \vec{v}&=&ct\vec{a}  +(\vec{\omega} \times
\vec{x}) \cr }.
\end{equation}

If one wishes to find the field corresponding to a given trajectory,
say, a geodesic in $CP(N-1)$, then, taking into account that any
geodesic as whole belongs to some $CP(1)$, one may put $ \pi^1=
e^{i\phi} \tan(\sigma \tau)$. Then $V^1=\frac{d \pi^1}{d
\tau}=\sigma \sec^2(\sigma \tau) e^{i\phi}$, and one has a linear
wave equations for the gauge unitary field $\Omega^{\alpha}$ in the
dynamical space-time with complicated coefficient functions of the
local coordinates $(\pi^1,...,\pi^{N-1})$.  Under the assumption
$\tau = w t$ this equation has following solution
\begin{eqnarray}
\omega+i \gamma=(F_1(r^2-c^2t^2)+i F_2(r^2-c^2t^2)) \exp{(-2w c
\int_0^t dp \frac{ \tan(w p)}{A \sqrt{c^2(p^2-t^2)+r^2}})},
\end{eqnarray}
where $F_1,F_2$ are an arbitrary function of the interval
$s^2=r^2-c^2t^2$, $(\vec{a},\vec{x})=Ar\cos(\chi)$,
$A=\sqrt{a_x^2+a_y^2+a_z^2}$ and $r=\sqrt{x^2+y^2+z^2}$. The angle
$\chi$ in fact is defined by a solution of the equation (5.20). I
used $\chi=\pi$ since for us now interesting only ``radial boost
turned toward the center of the field shell".

The general factor demonstrates the diffusion of the light cone
(mass shell) due to the boosts. Thus our results consist with the
so-called ``off-shell'' idea of Horwitz-Piron-Stueckelberg
\cite{H2}.

\section{Quasi-Hamiltonian equations}
 The theory of the quasi-liner PDE field equations
(\ref{FSE}) is well known \cite{Kamke,Arnold}. I wish apply general
approach looking on the quasi-linear PDE as a particular case of the
nonlinear PDE. Let me write such equation in the form
\begin{equation}\label{FSE1}
G(x^{\mu}, P_{\mu},\Omega)=v^{\mu} \frac{\partial \Omega}{\partial
x^{\mu} } + \Omega\Gamma^m_{mn} V^n=0,
\end{equation}
where I put $P_{\mu}=\frac{\partial \Omega}{\partial x^{\mu} }$.
Such PDE equations demonstrate a natural ``wave-corpuscular duality"
in the following sense. Equation (\ref{FSE1}) itself is field
equation relative $\Omega$. Then ``4-velocity" $v^{\mu} = \frac{d
x^{\mu}}{d \tau}$ against the parameter of evolution $\tau$ may be
treated as velocities of ``corpuscules" moving along trajectories in
the dynamical space-time. We can see it from the following
calculations. The equation (\ref{FSE1}) define the hyper-surface $E$
in the 9-dimension space of 1-jets \cite{Arnold}. The phase curves
lie as whole on $E$ and obey following quasi-Hamiltonian system.
\begin{equation}\label{FSE2}
\frac{d G(x^{\mu}, P_{\mu},\Omega)}{d \tau}=
G_{x^{\mu}}v^{\mu}+G_{P_{\mu}}\frac{d P_{\mu}}{d
\tau}+G_{\Omega}\frac{d\Omega}{d\tau}=0,
\end{equation}
Using the explicit form of $G$, one can rewrite this equation as
follows:
\begin{equation}\label{FSE3}
G_{x^{\mu}}v^{\mu}+v^{\mu} \frac{d P_{\mu}}{d
\tau}+G_{\Omega}\frac{\partial \Omega}{\partial x^{\mu}} v^{\mu}=0.
\end{equation}
Then the full characteristic system reads now
\begin{eqnarray}\label{FSE4}
\frac{d x^{\mu}}{d \tau} = \frac{\partial G}{\partial P_{\mu}} =
v^{\mu}, \cr \frac{d P_{\mu}}{d \tau}=-\frac{\partial G}{\partial
x^{\mu}} - \frac{\partial G}{\partial \Omega}P_{\mu}, \cr
\frac{d\Omega}{d\tau}=\frac{\partial \Omega}{\partial
x^{\mu}}v^{\mu}=\frac{\partial \Omega}{\partial
x^{\mu}}\frac{\partial G}{\partial P_{\mu}}=P_{\mu} \frac{\partial
G}{\partial P_{\mu}}.
\end{eqnarray}
Here one has generalized Hamiltonian equations describing a
``corpuscular" point-wise motion in the 4D dynamical space-time. The
analysis of stability of their solutions and the physical sense of
their Schr\"odinger's quantization require future investigations.
\section{Conclusion}
The main new points of my approach are following:

A. I use the notion of ``elementary quantum motions" (EQM) $|\hbar
a>$ with well defined quantized Planck's action $S_a=\hbar a$
instead of the notion of ``elementary particles". Their GCS's serve
as an abstract formalization of the ``quasi-classical'' description
of a quantum setup or ``Schr\"odinger's lump" \cite{Penrose}.

B. The quantum phase space $CP(N-1)$ serves as the base of the
tangent fibre bundle of the local dynamical variables. The special
cross-section of this bundle and affine gauge field are geometric
tools for the quantum measurement in the state-dependent dynamical
space-time.

C. Integration over all pathes (alternatives) realizes the objective
approach in quantum theory. The dominate contribution will be given
by the geodesic of $CP(N-1)$ spanning two GCS's \cite{Le4}.

The technical details are as follows:

1. The projective representation of pure $N$-dimension quantum
states (one could think of arbitrary large $N$), provides a natural
non-linear realization of the $G=SU(N)$ group manifold and the coset
sub-manifold $G/H=SU(N)/S[U(1)\times U(N-1)]=CP(N-1)$. I consider
the generators of this group as LDV's \cite{Le4} of the model.

2. These quantum dynamical variables are represented by the tangent
vector fields to $CP(N-1)$. Embedding of $CP(N-1)$ into
$\mathcal{H}=C^N$ provides the measurement procedure for the
dynamical variables.

3. Quantum measurement ``creates" local dynamical space-time capable
of detecting the coincidence of expectation and measured values of
these quantum dynamical variables.

4. The affine parallel transport, associated with the Fubini-Study
metric, accompanied with ``Lorentz spin transformation matrix"
\cite{G}, establish this coincidence due to the identification of
the parallel transported LDV at different GCS's.

5. The parametrization of the measurement results with the help of
{\it attributed} local space-time coordinates is in fact the
embedding of quantum dynamics in Hilbert space into 4D world. This
procedure is well definite due to the existence of the infinitesimal
$SL(2,C)$ transformations of the qubit spinor treated as Lorentz
transformations of local space-time coordinates.

6. The quasi-linear PDE for non-Abelian gauge field  in dynamical
space-time naturally related to the ODE of their characteristics.
The last ones are similar to the Hamilton canonical equations. Their
quantization leads to Schro\"odinger-like equations whose properties
will be discussed elsewhere.

\vskip 0.2cm ACKNOWLEDGEMENTS

I am sincerely grateful Larry Horwitz for interesting discussions
and critical notes.

\end{document}